\documentstyle[preprint,tighten,aps,prc]{revtex} 
\begin{document} 
\begin{titlepage} 
 
\begin{center} 
 
{\large{\bf $\pi \Delta\Delta$ coupling constant}} 
 
\vspace{1.2cm} 
 
Shi-Lin Zhu\\ 
2152 Hillside Road,  
Department of Physics, University of Connecticut, U-3046\\ 
Storrs, CT 06269-3046 
 
\end{center} 
 
\begin{abstract} 
 
We calculate the $\pi \Delta\Delta$ coupling $g_{\pi^0\Delta^{++}\Delta^{++}}$ using light 
cone QCD sum rule. Our result is $g_{\pi^0\Delta^{++}\Delta^{++}}=(11.8\pm 2.0)$. 
  
\vskip 0.5 true cm 
 
 
\end{abstract} 
\vspace{2cm} 
\vfill 
\end{titlepage} 
 
\pagenumbering{arabic}

The $\pi \Delta\Delta$ coupling constant $g_1$, like the nucleon axial charge $g_A$, is a basic 
parameter which enters the loop calculation in all processes involved with delta resonance 
\cite{barry} in chiral perturbation theory. Unfortunately it is not directly accessible experimentally. 
A special quartet scheme of chiral  
symmetry realization for even- and odd-parity baryon resonances
was proposed in \cite{prl}. Based on such a  
scheme the authors found that the parity nonchanging couplings such as $\pi \Delta_{\pm} N^\ast_\pm, 
\pi \Delta_\pm\Delta_\pm$, and $\pi N^\ast_\pm N^\ast_\pm$ are forbidden
at the leading order \cite{prl}. 
Such a result is very different from  
quark model prediction \cite{quark} and 
large $N_c$ argument $g_1={9\over 5}g_A$ \cite{nc}. 
Recently an attempt was made to extract this important coupling 
from the fit to the  
phase shift data of pion-nucleon scattering 
in the fourth order chiral perturbation theory analysis \cite{mei}. 
Because this coupling only appears in the third order loop 
contribution, it's very hard to pin down the value precisely. 
However the preliminary result was $g_1 =-0.94 \sim -2.65$ \cite{mei}.   
These value for $g_1$ comes out very differently 
from the large $N_c$ prediction
as noted in \cite{mei}.
So an independent theoretical extraction may prove useful to 
help clarify the present ambiguous situation concerning this coupling.  
 
We have calculated $\pi NN$ and $\pi N N^\ast$ \cite{zhu}, $\eta NN$ \cite{zhu-eta} 
and $\rho NN$, $\omega NN$ 
\cite{zhu1} coupling  constants in the framework of light cone QCD sum rule (LCQSR). The  
extracted values of various couplings from LCQSR are in good agreement with those used in or obtained  
from phenomenological analysis. In this short note we extend the same formalism to  
calculate the $\pi \Delta\Delta$ coupling constant.  
 
Let's first introduce some notations.  
For the $\Delta$ resonance, we use the isospurion formalism, treating the $\Delta$ field $T_\mu^i(x) $ 
as a vector spinor in both spin and isospin space with the constraint $\tau^i T_\mu^i 
(x)=0$ \cite{barry}. The components of this field are 
\begin{equation} 
T^3_\mu =-\sqrt{{2\over 3}}\left( \begin{array}{l} \Delta^+\\ \Delta^0 
 \end{array} \right)_\mu\ \ \ , \ T^+_\mu =\left( \begin{array}{l} 
\Delta^{++}\\ 
\Delta^+/\sqrt{3} 
 \end{array} \right)_\mu \ \ \ , \ T^-_\mu =-\left( \begin{array}{l} 
\Delta^0/\sqrt{3}\\ 
\Delta^- 
 \end{array} \right)_\mu\ \ \ . 
\end{equation} 
The field $T^i_\mu$ also satisfies the constraints for the 
ordinary Schwinger-Rarita spin-${3\over 2}$ field, 
\begin{equation} 
\gamma^\mu T_\mu^i=0\ \ \  {\hbox{and}}\ \ \  p^\mu T_\mu^i=0\ \ \ . 
\end{equation} 
To be specific, for the $\pi \Delta\Delta$ and $\pi NN$ interaction we use pseudoscalalar form: 
\begin{equation} 
{\cal L}_{PS} = g_{\pi\Delta\Delta} {\bar T}^\mu_i  i\gamma_5 {\vec \tau}\cdot {\vec \pi} T_\mu^i 
+g_{\pi NN} \bar N i\gamma_5 {\vec \tau}\cdot {\vec \pi} N +\cdots 
\end{equation} 
The pseudoscalar coupling $g_{\pi\Delta\Delta}$ can be realated to the pseudovector one $g_1$ 
via the relation:  
\begin{equation} 
g_1={F_\pi g_{\pi\Delta\Delta} \over m_\Delta} 
\end{equation} 
where $F_\pi =92.4$ MeV is the pion decay constant.  
 
Since the light cone QCD sum rule \cite{lcqsr} has proven useful in extracting  
strong coupling constants, we use it to calculate $g_{\pi\Delta\Delta}$ and consider 
the correlator: 
\begin{equation}\label{cor} 
i\int dx e^{ipx} <0|T\{ \eta_\mu (x), \bar \eta_\nu(0)\}|\pi^0 (q)> 
\end{equation} 
with the interpolating current for $\Delta^{++}$, $\eta_\mu (x)=\epsilon^{abc} [u^{aT}C\gamma_\mu 
u^b]u^c (x)$. We also introduce $<0|\eta_\mu(0)|\Delta^{++}>=\lambda_\Delta \nu_\mu$ with  
$\nu_\mu$ a vectorial spinor for the spin ${3\over 2}$ delta field and
$\lambda_{\Delta}$ is the overlapping amplitude. 
At the phenomenological side we consider the Lorentz structure 
$ig_{\mu\nu}{\not q}\gamma_5$ which admits contribution from resonances 
with $I=3/2, J=3/2$ only. 

The calculation is routine. We first make operator product expansion and
express Eq. (\ref{cor}) with pion light cone wave functions \cite{lcqsr}. 
After finishing Fourier tranformation we make double Borel transformation
twice to extract the double spectral density $\rho (s_1, s_2)$.
Finally we subract the continuum contribution to Eq. (\ref{cor}).
The present sum rule is symmetric with $s_1, s_2$ which enables 
a clean subtraction of the continuum. 
Similar and detailed calculation steps can be found in \cite{lcqsr,zhu}.
We present final sum rule directly.   
\begin{eqnarray}\label{cou-d}\nonumber 
m_\Delta \lambda_\Delta^2 g_{\pi^0\Delta^{++}\Delta^{++}}e^{-{m_\Delta^2\over M^2}} 
={F_\pi\over \pi^2}\{ {11\over 48}\varphi_\pi({1\over 2})M^6 f_2({s_\Delta\over M^2}) 
-{3\over 2}[g_1({1\over 2})  
+G_2({1\over 2})]M^4 f_1({s_\Delta\over M^2}) &\\ 
+{1\over 18}a \mu_\pi \varphi_\sigma({1\over 2}) M^2 f_0 ({s_\Delta\over M^2})\}+\cdots 
\end{eqnarray} 
where $M^2$ is the Borel parameter, $a=-4\pi^2 
\langle \bar q q\rangle =0.55$ GeV$^3$ is the quark condensate, 
$\mu_\pi=1.65$ GeV at the scale of 1 GeV, 
$s_\Delta$ is the continumm thresold of delta mass sum rule, 
$\varphi_\pi({1\over 2})$  
etc are the values of various pion wave functions at the point ${1\over 2}$.  
Their values are $\varphi_\pi({1\over 2})=1.5\pm 0.2, g_1({1\over 2}) +G_2({1\over 2}) 
=0.042, \varphi_\sigma({1\over 2})=1.47$ at the scale 1 GeV \cite{zhu,lcqsr}. 
Functions $f_n (x)=1-e^{-x}\sum\limits_{k=0}^{n}{x^k\over k!}$ are used to subtract the continuum 
contribution. We need delta mass sum rule \cite{hwang}: 
\begin{equation}\label{mass-d} 
(2\pi)^4\lambda_\Delta^2 e^{-{m_\Delta^2\over M^2}}={1\over 5}M^6 f_2({s_\Delta\over M^2}) 
-{5\over 72}b M^2 f_0 ({s_\Delta\over M^2}) +{4\over 3}a^2-{7\over 9}{a^2m_0^2\over M^2} 
\end{equation} 
where $b=\langle g_s^2 G^2\rangle =0.48$ GeV$^4$ is the gluon condensate, 
$m_0^2={\langle \bar q g_s\sigma\cdot G q\rangle\over 
\langle \bar q q\rangle}=0.8$ GeV$^2$. 

For comparison we collect the sum rule for $g_{\pi^0 PP}$  
and nucleon mass sum rule in literature below. 
\begin{eqnarray}\label{cou-n}\nonumber 
m_N \lambda_N^2 g_{\pi^0 PP}e^{-{m_N^2\over M^2}} 
={F_\pi\over \pi^2}\{ {1\over 2}\varphi_\pi({1\over 2})M^6 f_2({s_N\over M^2}) 
-4[g_1({1\over 2})  
+G_2({1\over 2})]M^4 f_1({s_N\over M^2}) &\\ 
+{1\over 9}a \mu_\pi \varphi_\sigma({1\over 2}) M^2 f_0 ({s_N\over M^2})\}+\cdots 
\end{eqnarray}  
\begin{equation}\label{mass-n} 
(2\pi)^4\lambda_N^2 e^{-{m_N^2\over M^2}}={1\over 2}M^6 f_2({s_N\over M^2}) 
+{1\over 8}b M^2 f_0 ({s_N\over M^2}) +{2\over 3}a^2-{1\over 6}{a^2m_0^2\over M^2} 
\end{equation}  
Eqs. (\ref{cou-d}) and (\ref{cou-n}) are very similar to each other.  
From our sum rule we do not expect 
$g_{\pi^0\Delta^{++}\Delta^{++}}$ to vanish while $g_{\pi^0 PP}$ remains large. 
This coupling can be extracted by dividing (\ref{cou-d}) by (\ref{mass-d}).  
The variation of $g_{\pi^0\Delta^{++}\Delta^{++}}$ with $M^2, s_\Delta$  
is shown in FIG. 1. The working region of the Borel parameter $M^2$
for the nucleon and delta sum rule is $1.0-1.5$ GeV$^2$ and
$1.2-1.7$ GeV$^2$ respectively \cite{hwang}. The upper three curves are 
for $g_{\pi^0 PP}$. The lower ones are for $g_{\pi^0\Delta^{++}\Delta^{++}}$.
From top to bottom the nucleon and delta continuum threshold takes
value of $s_N=2.35, 2.25, 2.15$ GeV$^2$, $s_\Delta=3.6, 3.5, 3.4$ GeV$^2$
respectively.

Numerically we get 
\begin{eqnarray}\nonumber 
g_{\pi^0\Delta^{++}\Delta^{++}}=(11.8\pm 2.0)\\ 
g_{\pi^0 PP}=(13.2\pm 1.5) 
\end{eqnarray} 
The central value corresponds to $M^2=1.4$ GeV$^2$, $s_N=2.25$ GeV$^2$,
$s_\Delta =3.5$ GeV$^2$.
The errors arise from the variation with the continuum threshold and
Borel parameter in the working region of the sum rules only.
Our calculation shows that the $\pi\Delta\Delta$ coupling is large 
although it is only half of the quark model 
\cite{quark}, SU(6) \cite{su6}, U(12) \cite{u12} and especially, 
large $N_c$ prediction \cite{nc}: 
$g_{\pi^0\Delta^{++}\Delta^{++}}= {9\over 5} g_{\pi^0 PP}$. 
It's interesting to note that our result is consistent with the  
phenomenological value extracted from an old isobar production 
experiment in $\pi^-p\to \pi^+\pi^-n$ 
near threshold \cite{mit}.

\vspace{2cm}
 {\bf Figure Captions}  
  
\begin{center}  
{\sf FIG 1.} {The variation of $g_{\pi^0\Delta^{++}\Delta^{++}}$
[$g_{\pi^0 PP}$] with the Borel parameter $M^2$ and the continuum threshold
$s_\Delta$ [$s_N$]. The upper and lower three curves are for
$g_{\pi^0 PP}$ and $g_{\pi^0\Delta^{++}\Delta^{++}}$ respectively.
From top to bottom $s_N=2.35, 2.25, 2.15$ GeV$^2$ and 
$s_\Delta=3.6, 3.5, 3.4$ GeV$^2$ respectively.}  
\end{center}  

\begin{thebibliography}{99} 
\bibitem{barry}T. R. Hemmert, B. R. Holstein and J. Kambor, J. Phys. G 24, 1831 (1998). 
 
\bibitem{prl}D. Jido, T. Hatsuda and T. Kunihiro, Phys. Rev. Lett. 84, 3252 (2000). 
 
\bibitem{quark}G. E. Brown and W. Weise, Phys. Rep. 22, 279 (1975). 

\bibitem{mei}N. Fettes and Ulf-G. Meissner, hep-ph/0006299. 
  
\bibitem{nc}R. Dashen, E. Jenkins and A. V. Manohar, Phys. Rev. D 49, 4713 (1994). 
 
\bibitem{zhu}Shi-Lin Zhu, W.-Y. P. Hwang and Yuan-Ben Dai, Phys. Rev. C 59, 442 (1999). 

\bibitem{zhu-eta}Shi-Lin Zhu, Phys. Rev. C 61, 065205 (2000). 
 
\bibitem{zhu1}Shi-Lin Zhu, Phys. Rev. C 59, 435 (1999); {\sl ibid.} C59, 3455 (1999). 
 
\bibitem{lcqsr}I. I. Balitsky, V. M. Braun and A. V. Kolesnichenko, Nucl. Phys. B 312, 509 (1989);\\ 
V. M. Barun and I. E. Filyanov, Z. Phys. C 48, 239 (1990);\\ 
V. M. Belyaev et al., Phys. Rev. D 51, 6177 (1995).  
 
\bibitem{hwang}W.-Y. P. Hwang and K.-C. Yang, Phys. Rev. D 49, 460 (1994). 
 
\bibitem{su6}F. Gursey, A. Pais and L. A. Radicati, Phys. Rev. Lett. 
13, 299 (1964). 
 
\bibitem{u12}B. Sakita and K. C. Wali, Phys. Rev. 139, 1355 (1965). 
 
\bibitem{mit}R. A. Arndt et al., Phys. Rev. D 20, 651 (1979). 
 
\end{thebibliography}
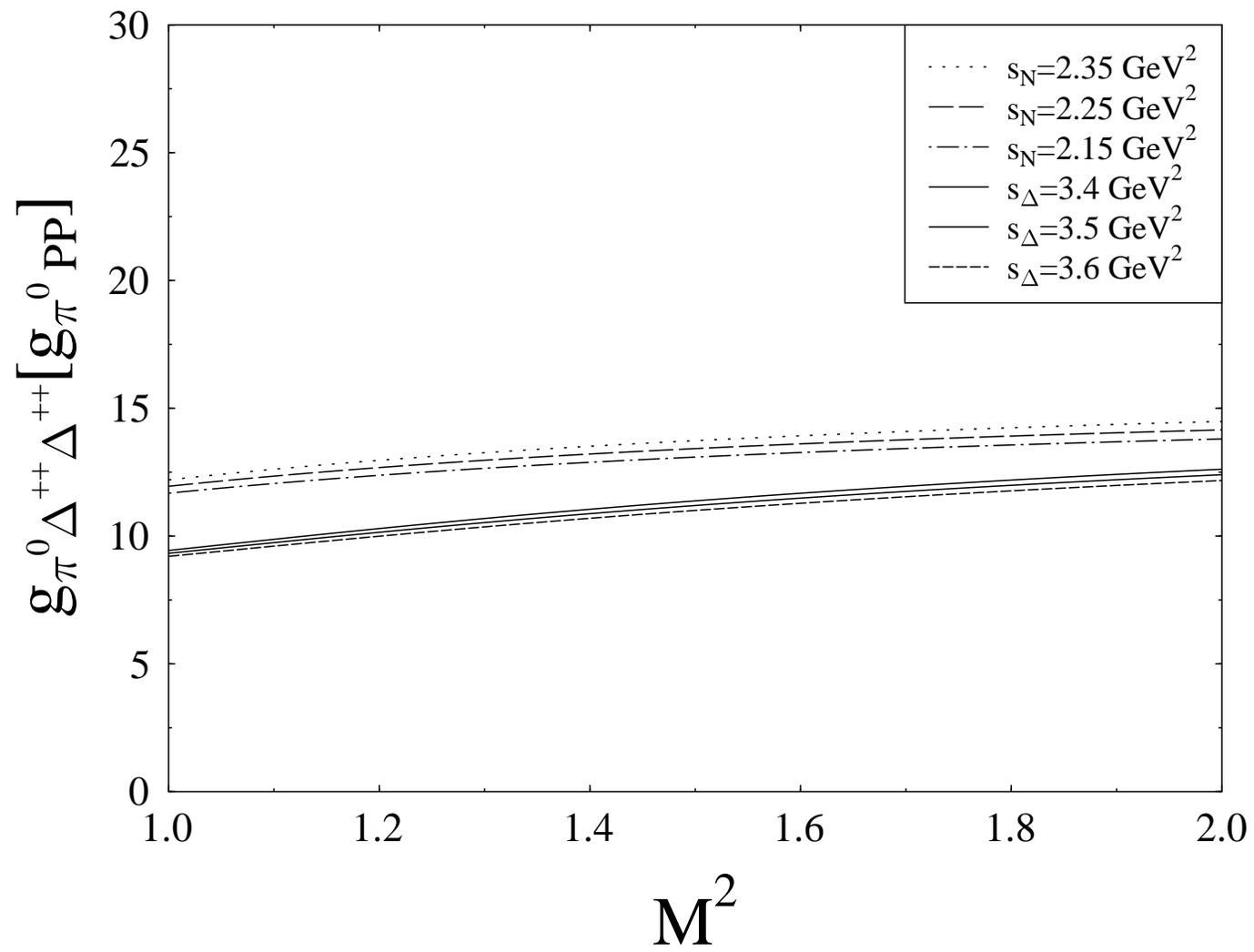
\end{document}